\documentclass[prd,superscriptaddress,twocolumn,amsmath,amssymb,nofootinbib,longbibliography]{revtex4-1}

\usepackage{graphicx}
\usepackage{amsmath}
\usepackage{physics}
\usepackage{color}
\usepackage[utf8]{inputenc}

\newcommand{\ba}{\begin{eqnarray}}
\newcommand{\ea}{\end{eqnarray}}

\newcommand{\eps}{\varepsilon}

\newcommand{\defterm}[1]{\emph{#1}}              %:text: %:ex: \defterm{definition}

\newcommand{\tens}[1]{{\ensuremath{\boldsymbol{#1}}}}       % please, use to make tensors bold
\newcommand{\ts}[1]{{\boldsymbol{#1}}}         % equivalent, just shorter

\renewcommand{\grad}{{\tens{d}}}                 % gradient                 %:ex: \grad f
\newcommand{\cv}[1]{{\tens{\partial}}_{#1}}    % coordinate vector        %:ex: \cv{t}
\newcommand{\pa}{\partial}                     % partial derivative

\newcommand{\A}[1]{A^{\!(#1)}}                 % metric functions         %:ex:  \A{j}_\mu
\newcommand{\dg}{{{n}}}                           % half-dimension
\newcommand{\KT}[1]{\tens{k}_{(#1)}}                  % KT       %:ex:  \KT{k}
\newcommand{\KVc}[1]{l_{(#1)}}                  % KV coordinates     %:ex:  \KVc{j}^{a}=\KTc{k}^{ac} \xi_{c}
\newcommand{\be}{\begin{equation}}             %:skip:
\newcommand{\ee}{\end{equation}}               %:skip:
\begin{document}

\title{Conformally Coupled Scalar in Rotating Black Hole Spacetimes}

%%% AUTHORS %%%<<<

\author{Finnian Gray}
\email{fgray@perimeterinstitute.ca}
\affiliation{Perimeter Institute, 31 Caroline Street North, Waterloo, ON, N2L 2Y5, Canada}
\affiliation{Department of Physics and Astronomy, University of Waterloo,
Waterloo, Ontario, Canada, N2L 3G1}

\author{Ian Holst}
\email{iholst@perimeterinstitute.ca}
\affiliation{Perimeter Institute, 31 Caroline Street North, Waterloo, ON, N2L 2Y5, Canada}
\affiliation{Department of Physics and Astronomy, University of Waterloo,
Waterloo, Ontario, Canada, N2L 3G1}

\author{David Kubiz\v n\'ak}
\email{dkubiznak@perimeterinstitute.ca}
\affiliation{Perimeter Institute, 31 Caroline Street North, Waterloo, ON, N2L 2Y5, Canada}
\affiliation{Department of Physics and Astronomy, University of Waterloo,
Waterloo, Ontario, Canada, N2L 3G1}

\author{Gloria Odak}
\email{godak@perimeterinstitute.ca}
\affiliation{Perimeter Institute, 31 Caroline Street North, Waterloo, ON, N2L 2Y5, Canada}
\affiliation{Department of Physics and Astronomy, University of Waterloo,
Waterloo, Ontario, Canada, N2L 3G1}

\author{Dalila M. P\^irvu}
\email{dpirvu@perimeterinstitute.ca}
\affiliation{Perimeter Institute, 31 Caroline Street North, Waterloo, ON, N2L 2Y5, Canada}
\affiliation{Department of Physics and Astronomy, University of Waterloo,
Waterloo, Ontario, Canada, N2L 3G1}
%>>>

\author{Tales Rick Perche}
\email{trickperche@perimeterinstitute.ca}
\affiliation{Perimeter Institute, 31 Caroline Street North, Waterloo, ON, N2L 2Y5, Canada}
\affiliation{Department of Physics and Astronomy, University of Waterloo,
Waterloo, Ontario, Canada, N2L 3G1}
\affiliation{Instituto de F\'{i}sica Te\'{o}rica, Universidade Estadual Paulista, S\~{a}o Paulo, S\~{a}o Paulo, 01140-070, Brazil}

\date{February 12, 2020}            % version 1.00; arxiv version 1

\begin{abstract}
We demonstrate separability of conformally coupled scalar field equation in general (off-shell) Kerr--NUT--AdS spacetimes in all dimensions.
The separability is intrinsically characterized by the existence of a complete set of mutually commuting conformal wave operators that can be constructed from a hidden symmetry of the principal Killing--Yano tensor. By token of conformal symmetry, the separability also works for any Weyl rescaled (off-shell) metrics. This is especially interesting in four dimensions where it guarantees separability of a conformally coupled scalar field in the general Pleba\'nski--Demia\'nski spacetime.
\end{abstract}

\maketitle
%%%%%%%%%%%%%%%%%%%%%%%%%%%%%%%%%%%%%%%

%%%%%%%%%%%%%%%%%%%%%%%%%%%%%%%%%%%%%%
%%%%%%%%%%%%%%%%%%%%%%%%%%%%%%%%%%%%
\section{Introduction}
The four-dimensional Kerr black hole geometry possesses many remarkable properties. Among these perhaps the most intriguing is the {\em separability} of various test field equations in the Kerr black hole background. The history of ``separatists'' began in 1968 when Carter demonstrated that the massive Hamilton--Jacobi and
Klein--Gordon equations can both be solved by a method of separation of variables \cite{Carter:1968rr, Carter:1968cmp}. Soon after that the massless wave equations for vector and tensor perturbations by Teukolsky \cite{Teukolsky:1972,Teukolsky:1973}, the massless neutrino equations by Unruh \cite{Unruh:1973} and Teukolsky \cite{Teukolsky:1973}, the massive Dirac equation by
Chandrasekhar \cite{Chandrasekhar:1976} and Page \cite{Page:1976}, and the Rarita--Schwinger equation by Kamran \cite{kamran1985separation} were all separated. Many of these results are directly linked to the existence of a hidden symmetry encoded in a Killing--Yano tensor \cite{Penrose:1973, Floyd:1973}. From this tensor one can generate other types of symmetries that enable a construction of symmetry operators underlying the separability of a given test field equation.

Interestingly, many of these results carry over to higher dimensions as well. In particular, the entire higher-dimensional Kerr--NUT--AdS family of vacuum black holes \cite{MyersPerry:1986, Gibbons:2004uw, Chen:2006xh} admits a hidden symmetry of the {\em principal Killing--Yano tensor}, a non-degenerate closed conformal Killing--Yano 2-form \cite{Kubiznak:2006kt}, which in its turn generates towers of explicit and hidden symmetries and implies separability of a number of test field equations in these backgrounds \cite{Frolov:2017kze}. Namely, the general separability of the massive Hamilton--Jacobi and scalar field equations was demonstrated in \cite{Frolov:2006pe}, the Dirac equation was separated in \cite{Oota:2007vx}, the massless and massive vector perturbations in \cite{Lunin:2017drx} and \cite{Frolov:2018ezx}, and the harmonic $p$-form equations in \cite{Lunin:2019pwz}, see also \cite{Oota:2008uj} for partial results on separability of gravitational perturbations.
In fact, all these separability results remain true in a more general class of spacetimes admitting the principal Killing--Yano tensor. The corresponding metric was  constructed in \cite{Houri:2007xz, Krtous:2008tb} and is known as the {\em off-shell Kerr--NUT--AdS spacetime}. Such a spacetime admits a number of unspecified metric functions and is not necessarily a solution to vacuum Einstein equations.

The purpose of the present paper is to extend the result on separability of the massive scalar equation demonstrated in \cite{Frolov:2006pe} and show that also the equation for a conformally coupled scalar,
\be\label{CCSF}
\bigl(\Box -\eta R\bigr)\Phi=0\,,\quad \eta=\frac{1}{4} \frac{d-2}{d-1}\,,
\ee
separates in the general off-shell Kerr--NUT--AdS spacetime. Here, $d$ stands for the number of spacetime dimensions, $R$ is the Ricci scalar of the background metric $\tens{g}$, and prefactor $\eta$ is chosen so that the equation enjoys conformal symmetry, see e.g. Appendix D in \cite{wald1984general}. As we shall demonstrate such a separability is intrinsically characterized by the existence of a complete set of commuting operators that are constructed from the tower of Killing tensors and Killing vectors generated from the principal Killing--Yano tensor of the Kerr--NUT--AdS geometry. Exploiting the conformal symmetry, the demonstrated separability remains valid for any Weyl rescaled metrics and in particular implies the separability of the conformal scalar equation in the general class of four-dimensional Pleba\'nski--Demia\'nski spacetimes.

Our paper is organized as follows: the general off-shell Kerr--NUT--AdS spacetime and its basic properties are introduced in Sec.~\ref{sec:2}. Separability of the conformal wave equation and its intrinsic characterization in this spacetime are studied in Sec.~\ref{sec:3}. The associated separability in Weyl scaled metrics is briefly discussed in Sec.~\ref{sec:3b}. Sec.~\ref{sec:4} is devoted to a concrete application of these results to four-dimensional spacetimes of the Pleba\'nski--Demia\'nski class. We summarize in Sec.~\ref{sec:5}.

%%%%%%%%%%%%%%%%%%%%%%%%%%%%%%%%%%%%%%%%%%%%%%%%%%
%%%%%%%%%%%%%%%%%%%%%%%%%%%%%%%%%%%%%%%%%%%%%%%%%%
\section{Kerr--NUT--AdS spacetimes}\label{sec:2}

The canonical metric describing the \defterm{off-shell Kerr--NUT--(A)dS geometry} in ${d=2n+\eps}$ number of dimensions (with $\eps=0$ in even and $\eps=1$ in odd dimensions) reads
\ba\label{KerrNUTAdSmetric}
\tens{g}=g_{ab}  \grad y^a \grad y^b&=&\sum_{\mu=1}^n\;\biggl[\; \frac{U_\mu}{X_\mu}\,{\grad x_{\mu}^{2}}
  +\, \frac{X_\mu}{U_\mu}\,\Bigl(\,\sum_{j=0}^{n-1} \A{j}_{\mu}\grad\psi_j \Bigr)^{\!2}
  \;\biggr]\nonumber\\
  && \qquad +\frac{\eps c}{\A{n}}\Bigl(\sum_{k=0}^n \A{k}\grad\psi_k\!\Bigr)^{\!2}\;.
\ea
The coordinates $y^a=\{x_\mu, \psi_k\}$ naturally split into two sets: \defterm{Killing coordinates} ${\psi_k}$ (${k}={0,\,\dots,\,\dg{-}1{+}\eps}$) associated with the explicit symmetries, and (Wick rotated) {\em radial and longitudinal coordinates} ${x_\mu}$  ($\mu=1,\,\dots,\,\dg$) labelling the orbits of Killing symmetries.
The functions ${\A{k}}$, ${\A{j}_\mu}$, and ${U_\mu}$ are `symmetric polynomials' of the coordinates ${x_\mu}$, and are defined by:
\begin{align}\label{AUdefs}
    \A{k}&=\!\!\!\!\!\sum_{\substack{\nu_1,\dots,\nu_k=1\\\nu_1<\dots<\nu_k}}^\dg\!\!\!\!\!x^2_{\nu_1}\dots x^2_{\nu_k}\;,\:\:\:\nonumber
\A{j}_{\mu}=\!\!\!\!\!\sum_{\substack{\nu_1,\dots,\nu_j=1\\\nu_1<\dots<\nu_j\\\nu_i\ne\mu}}^\dg\!\!\!\!\!x^2_{\nu_1}\dots x^2_{\nu_j}\;,\nonumber\\
U_{\mu}&=\prod_{\substack{\nu=1\\\nu\ne\mu}}^\dg(x_{\nu}^2-x_{\mu}^2)\;,\;\:\: U=\prod_{\substack{\mu,\nu=1\\\mu<\nu}}^n(x^2_\mu-x^2_\nu)=\det\bigl(\A{j}_\mu\bigr)\;,
\end{align}
where we have fixed $\A{0}=1=\A{0}_{\mu}$. Each metric function ${X_\mu}$ is an unspecified function of a single coordinate ${x_\mu}$:
\be
X_\mu=X_\mu(x_\mu)\,,
\ee
and finally the constant $c$ that appears in odd dimensions only is a free parameter.  We refer to \cite{Frolov:2017kze} for a prescription of how to translate this ``symmetric gauge'' to the Boyer--Lindquist type coordinates.

The inverse metric takes the form:
\ba\label{KerrNUTAdSinvmetric}
  \tens{g}^{-1}
  &=&\sum_{\mu=1}^n\!\left[\! \frac{X_\mu}{U_\mu}{\cv{x_{\mu}}^2}
  + \frac{U_\mu}{X_\mu}\!\left(\sum_{k=0}^{n-1+\eps}
    {\frac{(-x_{\mu}^2)^{n-1-k}}{U_{\mu}}}\!\cv{\psi_k}\!\right)^{\!\!2}\right]\nonumber\\
  &&+\eps\,\frac1{c\A{n}}\,\cv{\psi_n}^2\;,
\ea
while the square root of the determinant of the metric reads
\begin{equation}\label{detmetric}
    \sqrt{\abs{g}} = \bigl(c\A{n}\bigr)^\frac{\eps}{2}\, U\,.
\end{equation}
Despite the complexity of the metric the Ricci scalar (calculated in \cite{Hamamoto:2006zf}) is fairly simple given by a sum
\begin{equation}\label{RSeparated}
    R = \sum_{\mu = 1}^n \frac{r_\mu}{U_\mu}\,,
\end{equation}
of the functions
\be\label{rmu}
r_\mu=-X_\mu''-\frac{2\eps X_\mu'}{x_\mu}-\frac{2\eps c}{x_\mu^4}\,.
\ee
Importantly, each $r_\mu$ only depends on a single variable $x_\mu$.

The above off-shell Kerr--NUT--AdS spacetime is the most general metric that admits the principal Killing--Yano tensor \cite{Houri:2007xz, Krtous:2008tb}. This is a non-degenerate closed conformal Killing--Yano 2-form $\tens{h}$ obeying the equation
\be
\nabla_a h_{bc} = g_{ab}\,\xi_{c} - g_{ac}\,\xi_{b}\,,\quad \xi^a=\frac{1}{D-1}\nabla_bh^{ba}\,.
\ee
Explicitly, the principal Killing--Yano tensor is given by
\begin{equation}\label{PCCKYpot}
\tens{h}=\tens{db}\,,\quad \ts{b} = \frac12 \sum_{k=0}^{\dg-1}\A{k+1}\grad\psi_k\;.
\end{equation}
It generates towers of explicit and hidden symmetries, see \cite{Frolov:2017kze}. Namely, we obtain the following tower of closed conformal Killing--Yano tensors:
\be
\tens{h}^{(j)}=\frac{1}{j!}\underbrace{\tens{h}\wedge \dots \wedge \tens{h}}_{j\ \mbox{\tiny times}}\,.
\ee
Their Hodge duals are Killing--Yano tensors $\tens{f}^{(j)}=\tens{*h}^{(j)}$, and their square gives rise to a tower of rank-2 Killing tensors:
\be\label{f2}
k^{ab}_{(j)}=\frac{1}{(d-2j-1)!}f^{(j)a}{}_{c_1\dots d_{d-2j-1}}f^{(j)bc_1\dots c_{d-2j-1}}\,.
\ee
The latter take the following explicit form $(j=0, \dots, n-1)$:
\ba\label{KTjcoor}
  \KT{j}\!
  &=&\!\sum_{\mu=1}^n\! \A{j}_\mu\!\!\left[\! \frac{X_\mu}{U_\mu}\,{\cv{x_{\mu}}^2}\!
  + \!\frac{U_\mu}{X_\mu}\!\left(\!\sum_{k=0}^{n-1+\eps}\!
    {\frac{(-x_{\mu}^2)^{n-1-k}}{U_{\mu}}}\cv{\psi_k}\right)^{\!\!2}\!\right]\nonumber\\
  &&+\eps\,\frac{\A{j}}{c\A{n}}\cv{\psi_n}^2\,,
\ea
and generate the tower of Killing vectors:
\be\label{Killingcoord}
\boldsymbol{l}_{(j)} = \boldsymbol{k}_{(j)}\cdot \boldsymbol{\xi}=\cv{\psi_j}\,,
\ee
with an additional Killing vector in odd dimensions, $\boldsymbol{l}_{(n)}=\cv{\psi_n}$.
Note that the $j=0$ Killing tensor is just the inverse metric \eqref{KerrNUTAdSinvmetric}, and
the zeroth Killing vector is the primary Killing vector, $\boldsymbol{l}_{(0)}=\tens{\xi}=\cv{\psi_0}$.

Since all of the symmetries are generated by this single object $\tens{h}$, they all mutually Schouten--Nijenhuis commute (See \cite{Frolov:2017kze} for details):
\ba
\left[ \boldsymbol{l}_{(i)},\boldsymbol{k}_{(j)} \right]_{\mbox{\tiny SN}}&=&0\;,\; \left[ \boldsymbol{l}_{(i)},\boldsymbol{l}_{(j)} \right]_{\mbox{\tiny SN}}=0\;,\nonumber\\
\left[ \boldsymbol{k}^{(i)},\boldsymbol{k}^{(j)} \right]_{\mbox{\tiny SN}}&=& k^{(i)}_{e(a} \nabla^e k^{(j)}_{bc)} - k^{(j)}_{e(a} \nabla^e k^{(i)}_{bc)} = 0\;
\ea
The canonical coordinates $\{ x_\mu,\psi_k \}$ are thus completely determined by the principal Killing--Yano tensor: the coordinates $\psi_k$ are the Killing coordinates \eqref{Killingcoord} and the coordinates $x_\mu$ are the `eigenvalues' of $\tens{h}$, see \cite{Frolov:2017kze} for more details.

Let us finally mention that when the vacuum  Einstein equations are imposed, $G_{ab}+\Lambda g_{ab}=0$, the metric functions $X_\mu$ take the following explicit `polynomial' form:
\be
X_\mu=\sum_{k=\eps}^n c_k x_\mu^{2k}-2b_\mu x_\mu^{1-\eps}-\frac{\eps c}{x_\mu^2}\,,
\ee
where $c_n$ is related to the cosmological constant by $\Lambda=\frac{1}{2}(-1)^n(d-1)(d-2)c_n$, while other parameters $c_k, b_\mu, c$ are related to rotations, mass, and NUT charges. With these we recover the on-shell Kerr--NUT--AdS metrics constructed by Chen, L\"u, and Pope \cite{Chen:2006xh}.
However, in what follows we will not restrict to this specific case and we are going to work with the general off-shell Kerr--NUT--AdS metrics.

%%%%%%%%%%%%%%%%%%%%%%%%%%%%%%%%%%%%%%%%%%%%%%%%%%%%%%%%%%
%%%%%%%%%%%%%%%%%%%%%%%%%%%%%%%%%%%%%%%%%%%%%%%%%%%%%%%%%
\section{Separability of conformal wave equation}\label{sec:3}

%%%%%%%%%%%%%%%%%%%%%%%%%%%%%%%%%%%%%%%%%
\subsection{Conformal operators}
In order to separate the conformally coupled scalar field equation  \eqref{CCSF}, let us first consider the following operators:
\be
\tilde {\cal K}_{(j)} = \nabla_{a} k_{(j)}^{ab} \nabla_{b}\,,
\ee
whose explicit action on a scalar $\Phi$ reads
\be
\tilde {\cal K}_{(j)} \Phi=\nabla_{a} k_{(j)}^{ab} \nabla_{b} \Phi=\frac{1}{\sqrt{|g|}}\partial_a\Bigl(\sqrt{|g|}k^{ab}_{(j)}\partial_b \Phi\Bigr)\,.
\ee
To find the coordinate form of these operators, we use \eqref{KTjcoor} to obtain
\ba
\tilde {\cal K}_{(j)} \Phi&=&\sum_{\mu=1}^n \frac{1}{\sqrt{|g|}}\partial_\mu\Bigl(\sqrt{|g|}\frac{A_\mu^{(j)}X_\mu}{U_\mu} \partial_\mu \Phi\Bigr)\nonumber\\
&&+\sum_{\mu=1}^n \frac{A_\mu^{(j)}}{U_\mu X_\mu}\Bigl(\sum_{k=0}^{n-1+\eps}{(-x_\mu^2)^{n-1-k}}\partial_k\Bigr)^2\Phi\nonumber\\
&&+\eps \frac{A^{(j)}}{A^{(n)}}\partial_n^2 \Phi\,,
\ea
where we have abbreviated $\partial_\mu=\partial_{x_\mu}$, $\partial_k=\partial_{\psi_k}$, and $\partial_n=\partial_{\psi_n}$.
Employing the expression for the metric determinant \eqref{detmetric} and the fact that neither $A_\mu^{(j)}$ nor $U/U_\mu$ depend on coordinate $x_\mu$, we have
\ba
 \frac{1}{\sqrt{|g|}}\partial_\mu\Bigl(&&\sqrt{|g|}\frac{A_\mu^{(j)}X_\mu}{U_\mu} \partial_\mu \Phi\Bigr)\nonumber\\
 &&=\frac{A_\mu^{(j)}}{U_\mu}\frac{\partial_\mu\Bigl(\bigl(c\A{n}\bigr)^\frac{\eps}{2}X_\mu\partial_\mu \Phi\Bigr)}{\bigl(c\A{n}\bigr)^\frac{\eps}{2}}\nonumber\\
 &&=\frac{A_\mu^{(j)}}{U_\mu}\Bigl[\partial_\mu\bigl(X_\mu\partial_\mu \Phi\bigr)+\eps\frac{X_\mu}{x_\mu} \partial_\mu \Phi\Bigr]\,.
\ea
Finally using the following identity \cite{Frolov:2006pe}:
\be
\frac{A^{(j)}}{A^{(n)}}=\sum_{\mu=1}^n\frac{A_\mu^{(j)}}{U_\mu}\frac{1}{x_\mu^2}\,,
\ee
we arrive at the following explicit form of these operators:
\be
 \tilde {\cal K}_{(j)} \Phi= \sum_{\mu=1}^n\frac{\A{j}_\mu}{U_\mu}{ \tilde{\cal K}}_{(\mu)} \Phi\,, \label{Qopsplit}
\ee
where each $\tilde {\cal K}_{(\mu)}$ involves only one coordinate ${x_\mu}$ and reads
\ba
\tilde {\cal K}_{(\mu)}&=& \partial_\mu\bigl(X_\mu\partial_\mu \bigr)+
 \frac{1}{X_\mu}\Bigl(\sum_{k=0}^{n-1+\eps}{(-x_\mu^2)^{n-1-k}}\partial_k\Bigr)^2\nonumber\\
&&+\frac{\eps}{c x_\mu^2}\partial_n^2+\eps\frac{X_\mu}{x_\mu} \partial_\mu\,,
\ea
which is the form derived in \cite{Sergyeyev:2007gf}.

Let us next consider the following scalar functions
\be\label{Rj}
R_{(j)}=\sum_{\mu=1}^n \frac{A_\mu^{(j)}}{U_\mu} r_\mu\,,
\ee
where $r_\mu$ is given by Eq.~\eqref{rmu}, and upgrade the operators $\tilde {\cal K}_{(j)}$ above to the following ``{\em conformal operators}'':
\be\label{Kjupgraded}
{\cal K}_{(j)}=\tilde {\cal K}_{(j)}-\eta R_{(j)}\,.
\ee
We immediately find
\be
{\cal K}_{(j)} = \sum_{\mu=1}^n\frac{\A{j}_\mu}{U_\mu} {\cal K}_{(\mu)} \,,
\ee
where
\ba\label{Kmu}
{\cal K}_{(\mu)}&=& \partial_\mu\bigl(X_\mu\partial_\mu \bigr)+
 \frac{1}{X_\mu}\Bigl(\sum_{k=0}^{n-1+\eps}{(-x_\mu^2)^{n-1-k}}\partial_k\Bigr)^2\nonumber\\
&&-\eta r_\mu+\frac{\eps}{c x_\mu^2}\partial_n^2+\eps\frac{X_\mu}{x_\mu} \partial_\mu\,.
\ea

%%%%%%%%%%%%%%%%%%%%%%%%%%%%%%%%%%%%%%%%
\subsection{Separability}
Since $\tilde {\cal K}_{(0)}=\Box$, the conformally coupled scalar field equation \eqref{CCSF} can be written as
\be\label{our}
{\cal K}_{(0)}\Phi=0\,.
\ee
Slightly more generally, we can include a mass term and consider an equation
\be\label{our2}
\bigl({\cal K}_{(0)}-m^2\bigr)\Phi=0\,.
\ee
To separate this equation we seek the solution in the multiplicative separated form,
\begin{equation}\label{ansatz}
    \Phi = \prod_{\mu=1}^n Z_\mu(x_\mu)\prod_{k=0}^{n-1+\eps} e^{i\Psi_k \psi_k},
\end{equation}
where $\Psi_k$ are (Killing vector) separation constants and each of the $Z_\mu$ is a function of the single corresponding variable $x_\mu$ only.
With this ansatz we have
\be
    \partial_k\Phi = i\Psi_k\Phi\,,\quad  \partial_\mu \Phi = \frac{Z_\mu'}{Z_\mu}\Phi\,,\quad \partial^2_\mu\Phi = \frac{Z_\mu''}{Z_\mu}\Phi\,,
\ee
which allows us to rewrite \eqref{our} in the following form:
\begin{align}\label{GmuUmu}
    \Phi\sum_{\mu =1}^n\frac{G_\mu}{U_\mu}=0\,,
\end{align}
where $G_\mu=G_\mu(x_\mu)$ are function of one variable only
\ba
G_\mu&=& X_\mu\frac{Z_\mu''}{Z_\mu}+X_\mu' \frac{Z_\mu'}{Z_\mu}-\frac{1}{X_\mu}\Bigl(\sum_{k=0}^{n-1+\eps}{(-x_\mu^2)^{n-1-k}}\Psi_k\Bigr)^2\nonumber\\
&&-\eta r_\mu-\frac{\eps}{c x_\mu^2}\Psi_n^2+\eps\frac{X_\mu}{x_\mu} \frac{Z_\mu'}{Z_\mu}-m^2(-x_\mu^2)^{n-1}\,.
\ea
Here we have used another identity \cite{Frolov:2006pe}
\be
1=\sum_{\mu=1}^n\frac{(-x_\mu^2)^{n-1}}{U_\mu}\,.
\ee
Now let us use the following `{separability lemma}' \cite{Frolov:2006pe, Krtous:2007xg}
{\bf Lemma.}
{\em The most general solution of
\be
\sum_{\mu=1}^n \frac{f_\mu(x_\mu)}{U_\mu}=0\,,
\ee
where $U_\mu$ is defined in \eqref{AUdefs}, is given by
\be\label{sepfnu}
f_\mu=\sum_{k=1}^{n-1} C_k(-x_\mu^2)^{n-1-k}\,,
\ee
where $C_j$ are arbitrary (separation) constants.}

Thus, we see that the most general solution of \eqref{GmuUmu} is
\be\label{Gmumu}
G_\mu=\sum_{k=1}^{n-1} C_k(-x_\mu^2)^{n-1-k}\,.
\ee
That is, the equation \eqref{our2} is satisfied for our ansatz \eqref{ansatz} provided the
functions $Z_\mu=Z_\mu(x_\mu)$ satisfy the following ordinary differential equations (ODEs):
\ba\label{separated}
{Z_\mu''}&&+{Z_\mu'} \Bigl(\frac{X_\mu'}{X_\mu} +\frac{\eps}{x_\mu}\Bigr) -\frac{Z_\mu}{X_\mu^2}\Bigl(\sum_{k=0}^{n-1+\eps}{(-x_\mu^2)^{n-1-k}}\Psi_k\Bigr)^2\nonumber\\
&&-\frac{Z_\mu}{X_\mu}\Bigl(\eta r_\mu+\frac{\eps}{c x_\mu^2}\Psi_n^2+\sum_{k=0}^{n-1} C_k(-x_\mu^2)^{n-1-k}\Bigr)=0\,,\quad \ \
\ea
where we have set $C_{0}=m^2$. When the coefficient $\eta$ is set to zero, we recover the result from \cite{Frolov:2006pe} on separability of the massive Klein--Gordon equation in the off-shell Kerr--NUT--AdS spacetime in canonical coordinates.   On the other hand, setting $m=0$ we have successfully separated the conformal equation \eqref{CCSF} in these spacetimes.

%%%%%%%%%%%%%%%%%%%%%%%%%%%%%%%%%%%%%
\subsection{Commuting operators}
Following \cite{Sergyeyev:2007gf} let us now show that the above demonstrated separability can be `justified' by the existence of a complete set of mutually commuting operators. This set consists of the above constructed conformal operators ${\cal K}_{(j)}$ and the Killing vector operators ${\cal L}_{(j)}$,
\ba
{\cal K}_{(j)}&=&\nabla_{a} k_{(j)}^{ab} \nabla_{b}-\eta R_{(j)}\,,\label{Qopdef}\\
{\cal L}_{(j)} &=& - i\, \KVc{j}^{a} \nabla_{\!a}\,. \label{Lopdef}
\ea
To show that these operators mutually commute, we consider their explicit form
\ba
 {\cal L}_{(j)} &=& -i\,\frac{\pa}{\pa\psi_j}\;,\label{Lopcoor}\\
 {\cal K}_{(j)} &=& \sum_{\mu=1}^n\frac{\A{j}_\mu}{U_\mu}{ {\cal K}}_{(\mu)}\,, \label{CQopsplit}
\ea
where ${\cal K}_{(\mu)}$ were derived above and are given by equation \eqref{Kmu}.
Obviously, we have
\be
\label{opcomut}
       \bigl[ {\cal K}_{(k)},{\cal L}_{(l)}\bigr]=0\;,\quad
    \bigl[{\cal L}_{(k)},{\cal L}_{(l)}\bigr]=0\;.
\ee
To show that also
\be
\bigl[{\cal K}_{(k)},{\cal K}_{(l)}\bigr]=0
\ee
we can reuse the argument presented in \cite{Sergyeyev:2007gf}. First, note that for $\mu\neq\nu$ we have $[{\cal K}_{(\mu)}, {\cal K}_{(\nu)}]=0$ because these operators depend on different $x^\mu\neq x^\nu$ and so any derivatives terms will commute. Next we can employ the first of the following identities \cite{Frolov:2006pe}:
\be\label{inversions}
\sum_{k=0}^{n-1}A^{(k)}_\nu\frac{(-x_\mu^2)^{n-1-k}}{U_\mu}=\delta_{\mu}^{\nu}\,,\quad
\sum_{\mu=1}^n\frac{(-x_\mu^2)^{n-1-k}}{U_\mu} A_\mu^{(j)}=\delta^j_k\,,
\ee
to ``invert" the expression in \eqref{CQopsplit} to write
\be
{\cal K}_{(\mu)}=\sum_{k=0}^{n-1}(-x^2_\mu)^{n-1-k}{\cal K}_{(k)}\;.
\ee
Thus using $[{\cal K}_{(\mu)},(-x_\nu^2)^{n-1-l}]=0$ for $\mu\neq\nu$ we can express the commutation of the ${\cal K}_{(\mu)}$'s as
\be
0=[{\cal K}_{(\mu)},{\cal K}_{(\nu)}]=\sum_{k,l=0}^{n-1}(-x_\mu^2)^{n-1-k}(-x_\nu^2)^{n-1-l}[{\cal K}_{(l)},{\cal K}_{(k)}]\;.
\ee
In particular as the $(-x_\mu^2)^{n-1-k}$ are non-vanishing in general this shows that $[{\cal K}_{(k)},{\cal K}_{(l)}]=0$, as required.

Of course, the separated solution above is nothing else than the ``common eigenfunction'' of these operators and the separation constants $\{\Psi_k, C_j\}$ are the
corresponding eigenvalues, that is, for our solution \eqref{ansatz} obeying \eqref{separated} we have
\ba
{\cal K}_{(j)}\Phi&=&C_j \Phi\,,\\
{\cal L}_{(j}) \Phi&=& \Psi_j \Phi\,.
\ea
To see the former, we write
\ba
\frac{1}{\Phi}{\cal K}_{(j)}\Phi&=&\frac{1}{\Phi}\sum_{\mu=1}^n\frac{A_\mu^{(j)}}{U_\mu} {\cal K}_{(\mu)}\Phi\nonumber\\
&=&\sum_{\mu=1}^n\frac{A_\mu^{(j)}}{U_\mu} \Bigl(G_\mu+m^2(-x_\mu^2)^{n-1}\Bigr)\nonumber\\
&=&\sum_{\mu=1}^n\frac{A_\mu^{(j)}}{U_\mu}\sum_{k=0}^{n-1}C_k(-x_\mu^2)^{n-1-k}\nonumber\\
&=&\sum_{k=0}^{n-1}C_k\sum_{\mu=1}^n\frac{A_\mu^{(j)}}{U_\mu}(-x_\mu^2)^{n-1-k}=C_j\,,
\ea
where we have subsequently used \eqref{CQopsplit}, \eqref{Gmumu}, and the second identity \eqref{inversions}.

Now with the separability of the conformal wave equation guaranteed we can turn to applications involving metrics conformally related to general metric \eqref{KerrNUTAdSmetric}.

%%%%%%%%%%%%%%%%%%%%%%%%%%%%%%%%%%%%%%%%%%%%%%%%%%%%%%%%%%%%%%%%
%%%%%%%%%%%%%%%%%%%%%%%%%%%%%%%%%%%%%%%%%%%%%%%%%%%%%%%%%%%%%%%%
\section{Separability in Weyl rescaled metrics}\label{sec:3b}
The equation \eqref{CCSF} enjoys a conformal symmetry. This means that under a Weyl scaling of the metric,
\be
\tens{g} \to \ \tilde{\tens{g}}=\Omega^2\tens{g}\,,
\ee
we have \cite{wald1984general}
\be
\bigl(\tilde{\Box} -\eta \tilde R\bigr)\bigr[\Omega^{1-d/2}\Phi\bigr]=
\Omega^{-1-d/2}\bigl(\Box-\eta R\bigr)\Phi\,.
\ee
In other words, provided $\Phi$ is a solution to the equation \eqref{CCSF} in the spacetime with metric $\tens{g}$,
\be\label{phitilde}
\tilde{\Phi}=\Omega^{1-d/2} \Phi\;
\ee
is a solution of  \eqref{CCSF} in the spacetime with metric $\tilde{\tens{g}}$.

In particular, this implies that in any spacetime $\tilde{\tens{g}}$ related to the off-shell Kerr--NUT--AdS metric by the Weyl transformation,
we can find a solution of the corresponding conformal equation \eqref{CCSF} in the form \eqref{phitilde}, where $\Phi$ is the separated solution
\eqref{ansatz} and functions $Z_\mu$ obey \eqref{separated}. Strictly speaking, due to the pre-factor $\Omega^{1-d/2}$ the corresponding solution \eqref{phitilde} is no longer formally written in a multiplicative separation form and the corresponding separability is called {\em R-separability}.

Let us also note that this result is non-trivial as the principal tensor no longer exists in the Weyl scaled metrics and consequently only towers of conformal hidden symmetries (as opposed to full hidden symmetries) exist in the Weyl rescaled spacetimes. Specifically, if $\omega$ is a conformal Killing--Yano $p$-form in spacetime with $\tens{g}$, then $\tilde{\tens{\omega}}=\Omega^{p+1}\tens{\omega}$
is a conformal Killing--Yano $p$-form in spacetime with $\tilde{\tens{g}}$.
In particular
\be
\tilde{\tens{h}}=\Omega^3\tens{h}
\ee
is a new principal conformal Killing--Yano tensor, which however need no longer be closed and is a much weaker structure. This implies that each Killing tensor, generated from $j$ copies of $\tens{h}$ with $j+1$ contractions with the inverse metric, c.f. \eqref{f2}, becomes a conformal Killing tensor:
\be
\tilde{K}_{(j)}^{ab}={K}_{(j)}^{ab}\,,
\ee
and the former explicit symmetries become conformal Killing vectors, $\tilde{l}_{(j)}^a=l_{(j)}^a$.
It would be interesting to study how these symmetries can directly be applied to guarantee separability of conformal wave equations in these spacetimes.

%%%%%%%%%%%%%%%%%%%%%%%%%%%%%%%%%%%%%%
%%%%%%%%%%%%%%%%%%%%%%%%%%%%%%%%%%%%%%%%%
\section{Four-dimensional examples}\label{sec:4}

\subsection{Carter's spacetime}
To apply the above machinery, let us now specify to $d=4$ dimensions. Upon the Wick rotation of one of the $x_\mu$ coordinates,
\be
\psi_0=\tau\,,\quad \psi_1=\psi\,,\quad x_1=y\,,\quad x_2=ir\,,
\ee
and setting
\ba
X_1&=&-\Delta_y\,,\quad X_2=-\Delta_r\,,\nonumber\\
U_2&=&\Sigma=r^2+y^2=-U_1\,,
\ea
the off-shell Kerr--NUT--AdS spacetime yields the off-shell Lorentzian Carter's metric \cite{Carter:1968cmp},
\begin{align}
\label{Carter}
    \tens{g}=& -\frac{\Delta_{r}}{\Sigma}\left(\boldsymbol{d} \tau+y^{2} \boldsymbol{d} \psi\right)^{2}+\frac{\Delta_{y}}{\Sigma}\left(\boldsymbol{d} \tau-r^{2} \boldsymbol{d} \psi\right)^{2}  \\
    & +\frac{\Sigma}{\Delta_{r}} \boldsymbol{d} r^{2}+\frac{\Sigma}{\Delta_{y}} \boldsymbol{d} y^{2}\,, \nonumber
\end{align}
with arbitrary
\be
\Delta_r=\Delta_r(r)\,,\quad \Delta_y=\Delta_y(y)\,,
\ee
the principal  Killing--Yano tensor given by
\be
\tens{h}=y \grad y \wedge (\grad \tau -r^2 \grad \psi)-r\grad r \wedge (\grad \tau+y^2\grad \psi)\,,
\ee
and the following Ricci scalar:
\be
R=-\frac{\Delta_{r}''+\Delta_{y}''}{\Sigma}\,.
\ee

The conformal scalar field equation \eqref{CCSF} reduces to
\be\label{CCSF4d}
\bigl(\Box - \frac{R}{6} \bigr)\Phi = 0\,.
\ee
Its solution can be found in a separable form,
\be
\Phi= Z(r)\;Y(y)\;e^{i \omega \tau} e^{i \Psi \psi}\,,
\ee
where functions $Z$ and $Y$ satisfy the following ordinary differential equations:
\ba\label{ODE4d}
 \left( \Delta_r Z' \right)'&+&Z\Bigl(\frac{1}{\Delta_r}(\Psi + r^2 \omega)^2+\frac{\Delta_r''}{6}-C\Bigr)=0\,,\\
 \left( \Delta_y Y' \right)'&+&Y\Bigl(-\frac{1}{\Delta_y}(\Psi -y^2 \omega)^2+\frac{\Delta_y''}{6}+C\Bigr)=0\,.
 \ea

Of course, this result remains valid for the on-shell Carter spacetime \cite{Carter:1968cmp}, a solution to the Einstein--Maxwell--$\Lambda$ theory, for which
\ba
\Delta_{r}&=&\left(r^{2}+a^{2}\right)\left(1-\Lambda r^{2} / 3\right)-2 m r+e^2+g^2\,,\\
\Delta_{y}&=&\left(a^{2}-y^{2}\right)\left(1+\Lambda y^{2} / 3\right)+2 n y\,.
\ea
Here,  $e$ and $g$ are electric and magnetic charges, and $m, a, n$ are related to mass, rotation, and NUT charge parameters, while the metric is accompanied by the $U(1)$ gauge potential
\be\label{A}
\tens{A}=-\frac{er}{\Sigma}\bigl(\grad \tau + y^2\,\grad\psi\bigr)
-\frac{gy}{\Sigma}\bigl(\grad \tau - r^2\,\grad\psi\bigr)\;.
\ee

%%%%%%%%%%%%%%%%%%%%%%%%%%%%%%%%%%%%%%%%%%%%%%
\subsection{Pleba\'nski--Demia\'nski class}
Another, more general, class of 4-dimensional black hole spacetimes is encoded in the Pleba\'nski--Demia\'nski spacetime \cite{PlebanskiDemianski:1976}. The off-shell metric is given by
\be\label{PD}
\tilde{\tens{g}}=\Omega^2 \tens{g}\,,
\ee
where $\tens{g}$ is given in \eqref{Carter} and the conformal prefactor takes the following form:
\be
\Omega=\frac{1}{1 - yr}\,.
\ee
By the above theory, this spacetime admits a solution of the conformal equation \eqref{CCSF4d}, which can be found in the R-separated form
\be
\Phi= \frac{1}{\Omega}Z(r)Y(y)e^{i \omega \tau} e^{i \Psi \psi}\,,
\ee
where functions $Z$ and $Y$ obey the ordinary differential equations \eqref{ODE4d}.

One particular example of a spacetime in this class is the original on-shell Pleba\'nski--Demia\'nski metric \cite{PlebanskiDemianski:1976}, for which the metric functions $\Delta_r$ and $\Delta_y$ take the following specific
form:
\begin{align}
&\Delta_{r}=k+e^{2}+g^{2}-2 m r+\epsilon r^{2}-2 n r^{3}-(k+\Lambda / 3) r^{4},\\
&\Delta_{y}=k+2 n y-\epsilon y^{2}+2 m y^{3}-\left( k+e^{2}+g^{2}+\Lambda / 3\right) y^{4}\,,
\end{align}
where $e$, $g$, $n$, $k$, $m$, and $\epsilon$ are free parameters that are related to the electric and magnetic charges, NUT parameter, rotation, mass, and acceleration.
Due to the conformal invariance of Maxwell equations in $4d$, the gauge potential remains given by \eqref{A}. In this special case, the separability of the conformal scalar equation follows from the results presented in \cite{dudley1977separation}, see also \cite{kamran1985separation2} for its intrinsic characterization.

Another example of a spacetime which belongs to the off-shell Pleba\'nski--Demia\'nski class is the hairy black hole solution constructed in \cite{Charmousis:2009cm, Anabalon:2009qt}, see also \cite{Anabalon:2012ta} for a more general spacetime that can be written in the form \eqref{PD} with a more general conformal pre-factor.

%%%%%%%%%%%%%%%%%%%%%%%%%%%%%%%%%%%%%%
%%%%%%%%%%%%%%%%%%%%%%%%%%%%%%%%%%%%%%%%%
\section{Summary}\label{sec:5}

In this paper we have separated the conformal wave equation in general off-shell Kerr--NUT--AdS spacetimes in all dimensions, generalizing the work \cite{Frolov:2006pe} on separability of the massive Klein--Gordon equation in these spacetimes. Let us emphasize that although the two results formally coincide in vacuum with cosmological constant -- for the on-shell Kerr--NUT--AdS spacetime \cite{Chen:2006xh} -- they are very different for a more general matter content.

We have further shown that the demonstrated  separability can be intrinsically characterized by a complete set of mutually commuting operators. To the leading order in derivatives, these operators are constructed from Killing tensors and Killing vectors generated from the hidden symmetry of the off-shell Kerr--NUT--AdS spacetime encoded in the principal Killing--Yano tensor. The second order operators also pick up an ``anomalous'' absolute term, see \eqref{Rj} and \eqref{Kjupgraded}, which in the case of the original conformal wave operator is simply given by the Ricci scalar of the spacetime and guarantees the conformal invariance of the corresponding equation. It is plausible to conjecture that also for other operators ${\cal K}_{(j)}$ ($j=1,\dots, n-1$) these anomalous terms ensure some kind of conformal symmetry. Unfortunately, at the moment we only have a coordinate expression for these correction terms and cannot study conformal properties of these operators until a covariant expression for the  anomalous terms is found. This issue certainly deserves attention in the future.

We have also discussed the Weyl rescaled metrics and shown how our results imply separability of the conformal wave equations in those spacetimes. As a concrete application we have considered the most general type D spacetime described by the Pleba\'nski--Demia\'nski family and constructed the associated R-separated test field solution of the conformal wave equation. We expect that this construction applies to a wide class of solutions with various matter content, similar to what happens in four dimensions \cite{PlebanskiDemianski:1976, Charmousis:2009cm, Anabalon:2009qt, Anabalon:2012ta}.

The obtained separated solution \eqref{ansatz} is general -- it depends on $d-1$ separation constants $\{\Psi_k, C_j\}$ -- any solution to the conformal scalar equation can be written as a superposition of these separated modes. Note, however, that in our paper we have used the ``symmetric gauge'' \eqref{KerrNUTAdSmetric}, where the (Wick rotated) radial and longitudinal coordinates $x_\mu$ are treated on the same footing and there is no clear distinction between the time and angle Killing coordinates both being encoded in $\psi_k$. Consequently, also the resultant ordinary differential equations \eqref{separated} all ``look the same''. In order to apply our result to study the behavior of the scalar field in the black hole vicinity, one needs to transform to the `physical space', see \cite{Frolov:2017kze} where this is explicitly done. Upon this  one of the separated equations \eqref{separated} becomes a (distinguished) radial equation while the other equations are the angular ones.
In order to solve this system (which is only coupled through parameters of the solution and separation constants), one needs to impose the regularity conditions on the axes, as well as proper boundary conditions for the radial modes. This then distinguishes various physical modes one wants to study. For example, the quasinormal modes are characterized by ingoing boundary conditions on the horizon together with the appropriate asymptotic conditions. This in turn restricts the admissible values of the separation and integration constants, and poses the ``non-linear eigenvalue problem'', see e.g. \cite{Berti:2009kk, Dias:2015nua, Frolov:2018ezx} for how this is done in similar settings. In particular, we expect a similar approach to \cite{Frolov:2018ezx, Cayuso:2019ieu,Barragan-Amado:2020pad} (where comparable ODEs were obtained by exploiting the hidden symmetries for the case of massive vector fields) can be used to numerically analyze the quasinormal modes arising from the coupled ODEs \eqref{ODE4d} in the physically interesting Pleba\'nski--Demia\'nski family of spacetimes.

%%%%%%%%%%%%%%%%%%%%%%%%%%%%%%
\section*{Acknowledgements}
\label{sc:acknowledgements}
We would like to thank Gang Xu and Dan Wohns for organizing the PSI Winter School where this project was mostly completed, with special thanks to Robert who kept smiles on our faces,
and the PSI program for facilitating this research.  D.P. is the 2019/20 recipient of the Emmy Noether Scholarship, a PSI Honorary Award supported by the Emmy Noether Circle to help increase the number of women in physics and mathematical physics at Perimeter. The work was supported in part by the Natural Sciences and Engineering Research Council of Canada.
Research at Perimeter Institute is supported in part by the Government of Canada through the Department of Innovation, Science and Economic Development Canada and by the Province of Ontario through the Ministry of Economic Development, Job Creation and Trade.
\vspace{-1cm}

%\bibliographystyle{JHEP}
%\bibliography{refs}

\providecommand{\href}[2]{#2}\begingroup\raggedright\endgroup

\end{document}